\documentclass[12pt]{iopart}
 \usepackage{iopams}

\usepackage[T1]{fontenc}
\usepackage[english]{babel}
\usepackage{graphicx}
\usepackage[latin9]{inputenc}
\usepackage{mathpazo}

\begin{document}

\title{A tensor theory of space-time as a strained material continuum}
\author{A Tartaglia and N Radicella}
\address{Dipartimento di Fisica del Politecnico and INFN section of Turin,\\
Corso Duca degli Abruzzi 24, I-10129 Torino, Italy}
\ead{angelo.tartaglia@polito.it, ninfa.radicella@polito.it}

\begin{abstract}
The classical theory of strain in material continua is reviewed and
generalized to space-time. Strain is attributed to "external" (matter/energy
fields) and intrinsic sources fixing the global symmetry of the universe
(defects in the continuum). A Lagrangian for space-time is worked out,
adding to the usual Hilbert term an "elastic" contribution from intrinsic
strain. This approach is equivalent to a peculiar tensor field, which is
indeed part of the metric tensor. The theory gives a configuration of
space-time accounting both for the initial inflation and for the late
acceleration. Considering also the contribution from matter the theory is
used to fit the luminosity data of type Ia supernovae, giving satisfactory
results. Finally the Newtonian limit of the theory is obtained.
\end{abstract}



\maketitle

\section{Introduction}

There are in cosmology a limited but important number of facts which do not
fit in the classical general relativistic (GR) frame of the standard theory.
Among others we recall the current accelerated expansion of the universe, as
discovered at the end of 1998 observing the redshift dependence of the
luminosity of type Ia supernovae \cite{perlmutter99,riess98}, the extreme
uniformity of the temperature of the Cosmic Microwave Background (CMB) \cite
{debernardis00,hanany00,spergel07}, the anomalous rotation curves of spiral
galaxies \cite{zwicky33,zwicky37}, the anomalous mass-luminosity relation in
galaxy clusters \cite{stanek} and the galaxy power spectra \cite{tegmark04}.
If, on one side, the number of facts to be accounted for is limited, on the
other, the number of theories put forth to explain them is extremely high.
All these theories may however be grouped under a limited number of
headings. One is GR plus appropriate fields. We may include in this category
the inflaton, accounting for the homogeneity of the CMB \cite
{guth,linde,lemoine}, as well as quintom \cite{linde94,cai}, k-essence \cite
{armendariz99,garriga} and vector inflation \cite{golovnev08}. Then we find
dark energy in a number of variants: cosmological constant \cite
{einstein17,einstein31,sahni,carrollreview}, quintessence\cite
{zlatev99,carroll98}, phantom energy \cite{caldwell02,vikman}, holographic
dark energy \cite{zhang,li,pavon}, triads \cite{armendariz} and k-essence
again \cite{armendariz00}. Together with the idea of dark energy we may
quote an, in principle, different ingredient of the universe which is dark
matter\cite{delpopolo}; now the competition opens among many possible types
of non-baryonic matter and specific candidates: Hot Dark Matter \cite
{hannestad08}, which actually seems not to be the right stuff since it does
not allow for structure formation; Cold Dark Matter \cite{dubovsky05}; Warm
Dark Matter \cite{palazzo07}; neutrinos, gravitinos, photinos, axions,
sterile neutrinos, Weakly Interacting Massive Particles (WIMPS), Lightest
Supersymmetric Particles (LSP). By the way the most successful cosmological
theory of the moment is the so called Lambda-Cold Dark Matter theory ($
\Lambda CDM$) \cite{LCDM}

Another compartment is the one of extended, alternative, modified theories
of gravity (all changes are with respect to standard GR). In this group we
find both metric and Palatini versions of $f\left( R\right) $ theories \cite
{capozziellofrancaviglia, faraonisotiriou} and Modified Newtonian Dynamics
(MOND) \cite{milgrom1,milgrom2,milgrom3} and its covariant formulation \cite
{bekenstein}.

All these possibilities remain within the realm of classical theories, but
of course we have also a host of attempts to introduce quantum aspects in a
more or less fundamental way. So we find various string cosmologies \cite%
{mavromatos02}, brane world \cite{maartensreview}, Loop Quantum Cosmology 
\cite{LQC1,bojowaldreview}, and similar approaches.

Everything can be good or not according to the capability of one or another
idea to account for a consistent enough number of facts, which, as said, are
rather few. The impression one has, however, is that most theories rely much
more on the mathematics of the model rather than on physics. In some cases
we have more or less explicit tricks introduced in order to produce the
right equations. The wording is in any case physical, but the entities that
are called in often have rather exotic properties, in the sense that they
are completely unusual, i.e. not corresponding to anything else we know. So,
considering the various entities that are introduced in the universe in
order to account for the accelerated expansion we have "fluids" which do not
dilute even though the available room is increasing (cosmological constant),
violating the strong energy condition (quintessence), such that when
compressed they expand (phantom energy). Mathematically everything is OK but
someone may feel uneasy with all that belonging to the actual world.

In order to find a route in the wild forest of conjectures one needs a
criterion and one could be to stick, as far as possible, to the physical
properties one already knows and has learned to describe. Of course there is
no reason why the universe on the largest scale should possess properties
that look like the ones we are used to at everyday's scale, but we have some
hints telling us that this approach could be not completely absurd. On one
side we know that the world becomes richer at the tiniest scales than at our
usual scale: wave functions and the whole machinery of quantum mechanics are
not needed to describe the world at the human scale. On another side we know
that some properties of complex systems are scale invariant. On this more or
less inspiring basis we have tried to draw on a known methodology and a
known theory in order to describe space-time as such. In GR space-time is
indeed described as a real continuum which interacts with something else,
being this "something else" what we call matter/energy; the interaction is expressed by the
non-linear Einstein equations whereby the non-linearity implies that
space-time, in a sense, interacts also with itself. Following on we
easily see that many techniques used for space-time in GR are very similar
to the ones applied to describe material continua in the theory of
elasticity: in the end, in both cases, everything is dominated by geometry.
How far this analogy can go? How the basic tools of the elasticity theory
need be modified in order to fit the behaviour of a four-dimensional
manifold with Lorentzian signature? The idea is not at all new and in a
sense has distant roots even in the old investigations on the nature of the
"luminiferous ether" that haunted physics until the beginning of the XXth
century. Of course the "ether" was thought of as something filling space and
moving or vibrating in an absolute time, the mathematical tools and concepts
were the ones of the XIX century and the aim of the research was the
explanation of the propagation of light. However, as we know, after getting
rid of the name "ether" Einstein remarked (1920) that "\textit{...space is
endowed with physical qualities; in this sense, therefore, there exists an
ether...}" \cite{Einstein}, and in today's quantum field theories the ether
is named "the vacuum" and has indeed important physical properties. So what
we are about to do is to explore the idea of space-time as a material
continuum using the formalism and general framework of GR and the leading
idea of the material continua we know. The central concept will be the
strain of the four-dimensional manifold which is related to the metric
tensor and is in turn produced by some cause. The origin of the strain can
be either "external" or internal and in our case, according to the
universally adopted dualistic approach, "external" will mean matter/energy.
The possibility of internal or spontaneous strain states is well known in
the theory of elasticity and plasticity and is connected with the presence
of structural defects in the "texture" of the manifold. So the other concept
we shall use in our theory will be the one of space-time defect or Cosmic
Defect (CD) which we use also to name the theory.

We have already applied this approach heuristically to the interpretation of
the late acceleration of the universe, obtaining encouraging results \cite%
{tartagliacapone,tartagliaaltri}. Here we are giving a new formulation of
the theory, casting it in a systematic and general way.

\section{Review and reformulation of the metric properties of elastic
continua}

\label{Review and reformulation of the metric properties of elastic continua}

For our purposes we shall consider an $N-$dimensional material continuum and
assume that in its reference state it is perfectly homogeneous and isotropic
so that we may identify it with an $N-$dimensional Euclidean manifold
(reference manifold) \cite{landau, eshelby}.

If we now imagine to apply some action (in three dimensions we would speak
of forces) either localized or diffused, we expect the continuum to be
brought to a deformed (strained) state, where in general it, as a manifold,
will acquire curvature: this will be our "natural" manifold \cite{eshelby}.

Let us then imagine we have two copies of the system, one in the reference
state, the other in the natural one, and that they are both embedded in an $%
N+1$ - dimensional simply-connected flat Minkowskian manifold \footnote{%
Actually for general natural manifolds it can happen that more than $1$
extra-dimensions are needed in order to build a flat embedding manifold \cite%
{rosen65}. In particular, it is known that an embedding of a non-vacuum
solution in five dimensions is minimal while an embedding of a non-flat
vacuum solution in six dimensions is minimal.}; the additional dimension will be time-like. We consider a situation in
which it is possible to put in one to one correspondence points on the two
manifolds. This operation may be formally thought of as a type II gauge
transformation \cite{sachs64,valsakumar}.

\begin{figure}[tbph]
\begin{center}
\includegraphics[width=10cm]{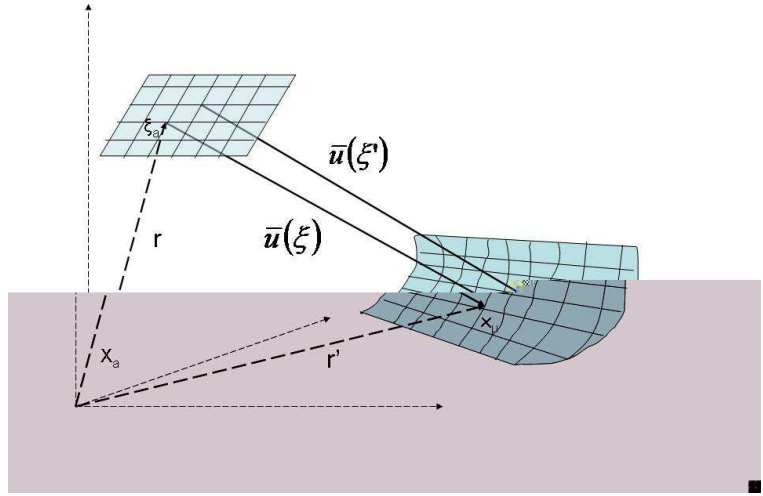}
\end{center}
\caption{Embedding of the reference and the natural manifolds in an $N+1$ -
dimensional flat manifold. $X_{a}$ are the Cartesian coordinates in the
embedding manifold; $\protect\xi _{\protect\mu }$ are the coordinates on the
reference (sub)-manifold ($N$ - dimensional), $x_{\protect\nu }$ are the
coordinates in the natural (generally curved) (sub)-manifold. $u$ represents
the displacement vector from points of the reference manifold to points of
the natural manifold.}
\label{fig00}
\end{figure}

Considering the embedding $N+1$ - manifold we may introduce a "displacement
vector field" $u$ which connects pairs of points in the two embedded $N$
-dimensional manifolds. If $r$ is the $N+1$- vector localizing a point in
the reference manifold, the $u$ vector runs from that position to a
different one $r^{\prime }$ in the natural manifold. The situation is shown
on fig.( \ref{fig00}). $X$'s are the (Cartesian) coordinates of the
embedding manifold; the coordinates on the reference manifold are $\xi$'s;
the ones of the natural manifold are $x$'s. We assume that the functional
dependences leading from one coordinate system to the other are all
sufficiently smooth for the rest of our reasoning to be valid. Of course the
different dimensionality implies the existence of at least one constraint
for each of the two sub-manifolds allowing for the dimensional reduction. In
general we assume $\xi _{\mu }=f_{\mu }\left( X_{1},...,X_{N+n}\right) $ and
(on the reference sub-manifold) $X_{i}=h_{i}\left( X_{1},...,X_{N}\right) $
for $i=N+1,...,N+n$ ; the same holds for the natural manifold: $x_{\alpha
}=w_{\alpha }\left( X_{1},...,X_{N+n}\right) $ and $X_{i}=q_{i}\left(
X_{1},...,X_{N}\right) $ for $i=N+1,...,N+n$. If all functions are
differentiable and invertible one can also write: 
\begin{equation}
x_{\mu }=m_{\mu }\left( \xi _{1},...,\xi _{N}\right)  \label{emme}
\end{equation}

On these bases we may write, with reference to the embedding manifold: 
\begin{equation}
r^{\prime }\left( X\right) =r\left( X\right) +u\left( X\right)
\label{vettori1}
\end{equation}

The functions $f$ and $w$ together with the constraints allow to cast (\ref%
{vettori1}) in terms of the reference $\{\xi \}$ or natural $\{x\}$
coordinates. A choice one can always do is to numerically identify $\xi $'s
and $x$'s for corresponding places in the two manifolds, not forgetting that
this choice can be practically useful but has per se no physical meaning.

Everything is non-trivial only in the case of $u$ being a non-trivial field:
rigid translations are uninteresting and amount to a simple coordinate
change in the same manifold.

The next step we may think to do is to compare the distances between pairs
of corresponding points in the unstretched and stretched situations. This
comparison is meaningful only if made in one and a single manifold, i.e. the
embedding one. Using unprimed symbols for the reference manifold and primed
symbols for the natural one we may write

$$
\begin{array}{l}
dl^{2}=\eta _{ab}dX^{a}dX^{b} \\ 
dl^{\prime 2}=\eta _{ab}dX^{\prime a}dX^{\prime b}
\end{array}
\quad \quad a,b =1,...,N+n
$$

In both cases the metric tensor is the one pertaining to the flat embedding
manifold. Introducing the constraints defining the two sub-manifolds of
interest and using the coordinates adopted for each of them, the equations above become 
$$\begin{array}{lll}
dl^{2}= & \eta _{ab}\frac{\partial X^{a}}{\partial \xi ^{\mu }}\frac{%
\partial X^{b}}{\partial \xi ^{\nu }}d\xi ^{\mu }d\xi ^{\nu }= & \eta
_{\alpha \beta }d\xi ^{\alpha }d\xi ^{\beta }=\mathsf{\eta }_{\mu \nu
}dx^{\mu }dx^{\nu } \\ 
dl^{\prime 2}= & \eta _{ab}\frac{\partial X^{\prime a}}{\partial x^{\mu }}%
\frac{dX^{\prime b}}{\partial x^{\nu }}dx^{\mu }dx^{\nu }= & g_{\mu \nu
}dx^{\mu }dx^{\nu } \\ 
\end{array}
 \qquad \quad \mu ,\nu =1,...,N
$$
Saturation of the Latin indices leads, on the first row, to the $N$
-dimensional flat metric tensor of the reference manifold (either Euclidean or
Minkowskian); however it is
important to stress that the additional factors $\partial \xi ^{\alpha
}/\partial x^{\mu }$ do not correspond to a coordinate change on the same
manifold, so that the final $\mathsf{\eta }_{\mu \nu }$ tensor is not in
general a metric tensor on the natural manifold. Viceversa on the second
line, the constraint to stay on the natural manifold leads in general to a
curved $N$ - dimensional metric (with Lorentzian signature). Using (\ref{vettori1}) and (\ref{emme}) we
see that it is: 
\begin{equation}
g_{\mu \nu }\left( x\right) =\mathsf{\eta }_{\mu \nu }+2\varepsilon _{\mu
\nu }\left( x\right)   \label{metrica01}
\end{equation}%
where $\varepsilon _{\mu \nu }$ is the strain tensor of the natural
manifold, given by 
\begin{equation}
\varepsilon _{\mu \nu }=\frac{1}{2}\left( \eta _{a\mu }\frac{\partial u^{a}}{%
\partial x^{\nu }}+\eta _{\nu b}\frac{\partial u^{b}}{\partial x^{\mu }}%
+\eta _{ab}\frac{\partial u^{a}}{\partial x^{\mu }}\frac{\partial u^{b}}{%
\partial x^{\nu }}\right)   \label{strain}
\end{equation}

Looking at (\ref{strain}) we easily see that the strain we have defined
transforms as a genuine tensor on the natural four-dimensional manifold.\\
If one could find a coordinate transformation such that (\ref{metrica01})
could be written as 
\begin{equation}
g_{\mu \nu }=\eta _{\alpha \beta }\frac{\partial \xi ^{\alpha }}{\partial
x^{\mu }}\frac{\partial \xi ^{\beta }}{\partial x^{\nu }},  \label{desaint}
\end{equation}%
then the strained and the unstrained situations would be diffeomorphic to
each other ($\mathsf{\eta }_{\mu \nu }$ would coincide with $g_{\mu \nu }$)
and it would be impossible to perceive the deformation from within the
manifold: no intrinsic curvature. In fact the integrability condition for (%
\ref{desaint}) is De Saint Venant's: 
\begin{equation}
R_{\beta \gamma \delta }^{\alpha }=0,  \label{riemann}
\end{equation}%
where $R_{\beta \gamma \delta }^{\alpha }$ are the components of the Riemann
tensor. This is indeed the case of elastic deformations, which have an
exogenous origin: they are due to the application of external forces and the
strain is brought to zero whenever those forces are removed (absence of
plastic deformations). In practice a smooth and continuous $u$ field leads
to an integrable (\ref{desaint}). If we are interested in intrinsic
deformations (the ones that can be sensed from within) we must study
singular $u$ (and $\varepsilon $) fields, the singularity being represented
by some discontinuity in $u$ and/or its derivatives. Now, a singular
displacement field in a continuum means that the medium contains what is
formally defined as one defect (or more), according to the definition given
by Volterra in 1907, while studying elastic and plastic deformations \cite%
{volterra}.

The singularity in $u$ reflects of course also in the strain tensor and in
general we shall write \cite{puntigam97,nabarro,hirth} the elementary
deformation (from (\ref{vettori1}) in the reference frame) as a non
integrable one form: 
\begin{equation*}
dr^{\prime \mu }=\omega _{\nu }^{\mu }dr^{\nu },
\end{equation*}
so that the deformed line element will be 
\begin{equation*}
ds^{2}=g_{\mu \nu }\left( x\right) dx^{\mu }dx^{\nu }=\eta _{\alpha \beta
}\omega _{\lambda }^{\alpha }\omega _{\rho }^{\beta }d\xi ^{\lambda }d\xi
^{\rho }
\end{equation*}

Now, our target is space-time and we know that interesting situations there
imply condition (\ref{riemann}) to be violated so that we are led to the
conclusion that any non-trivial space-time should contain at least one
defect in the sense recalled above. This is a different way to the
singularity theorems by Hawking and Penrose \cite{hawking73}.  A warning that
is appropriate to issue at this point is that the defect we are considering
here should not be confused with the topological ones often appearing in
cosmology. There the topological defects are the residual of the phase
transition which gave origin to matter during inflation. Of course "our"
defects also have topological properties, however their nature is completely
different, as we have seen.

\section{Lagrangians for space-time}

After the geometric considerations developed in the previous section we are
left with the problem of finding the functional dependence of $u$ (or to say
better, $\varepsilon $) on the coordinates. In practice this means that we
need to introduce an appropriate Lagrangian for our space-time manifold.

Let us start from the typical Einstein-Hilbert Lagrangian for a defectless
manifold 
\begin{equation}
L_{EH}=R\sqrt{-g},  \label{LHE}
\end{equation}
which is build from the simplest scalar obtainable from the curvature
tensor. We of course expect that the Lagrangian we are looking for reduces
to (\ref{LHE}) when the defect disappears. An apparently reasonable approach
is to add to (\ref{LHE}) an "elastic" potential term: this would be
consistent with the description given so far.

A typical expression for the elastic potential energy is: 
\begin{equation}
W_{e}=\frac{1}{2}\sigma _{\mu \nu }\varepsilon ^{\mu \nu }
\label{potenziale}
\end{equation}
where the usual meaning of $\sigma _{\mu \nu }$ is to be the components of
the stress tensor of the material. The stress is naturally dependent on the
strain and viceversa. In the standard elasticity theory this dependence is
mostly dealt with assuming a linear relation (Hooke's law) \cite{landau}. In
the case of space-time we have a priori no reason to say that it is so also,
but we shall assume linearity and see what happens. We write: 
\begin{equation*}
\sigma _{\mu \nu }=C_{\mu \nu \alpha \beta }\varepsilon ^{\alpha \beta }
\end{equation*}
where $C_{\alpha \beta \mu \nu }$ are the components of the elastic modulus
tensor and are supposed to be independent from the $\varepsilon $'s. As a
consequence (\ref{potenziale}) becomes: 
\begin{equation}
W_{e}=\frac{1}{2}C_{\mu \nu \alpha \beta }\varepsilon ^{\alpha \beta
}\varepsilon ^{\mu \nu }  \label{we}
\end{equation}

We have of course to do with a scalar density and we may study the situation
in a locally flat manifold (tangent space). Assuming that the medium, at
least in the unperturbed condition, is perfectly homogeneous and isotropic
the form assumed by the elastic modulus tensor depends on two parameters
only \cite{landau} and is 
\begin{equation}
C_{\alpha \beta \gamma \delta }=\lambda \eta _{\alpha \beta }\eta _{\gamma
\delta }+\mu \left( \eta _{\alpha \gamma }\eta _{\beta \delta }+\eta
_{\alpha \delta }\eta _{\beta \gamma }\right)  \label{moduli}
\end{equation}
The parameters $\lambda $ and $\mu $ are known as the Lamé coefficients;\
their value is a property of the continuum under consideration. In the
standard elasticity theory instead of the $\eta $'s one has Kronecker deltas
(at least for Cartesian coordinates), i.e. the Euclides metric tensor; here
we introduce the Minkowski tensor, because we are dealing with space-time.
In the case of space-time the global isotropy condition in four dimensions
has to be taken with caution, because of the light cones, however we shall
assume it holds. Considering (\ref{moduli}) and (\ref{we}) we arrive at 
\begin{equation*}
W_{e}=\left( \frac{1}{2}\lambda \varepsilon ^{2}+\mu \varepsilon _{\alpha
\beta }\varepsilon ^{\alpha \beta }\right)
\end{equation*}
being $\varepsilon =\varepsilon _{\alpha }^{\alpha }$ the trace of the
strain tensor. In the globally curved manifold the corresponding Lagrangian
density is: 
\begin{equation}  \label{elasto}
L_{e}=\left( \frac{1}{2}\lambda \varepsilon ^{2}+\mu \varepsilon _{\alpha
\beta }\varepsilon ^{\alpha \beta }\right) \sqrt{\left\vert g\right\vert }
\end{equation}
Of course indices are raised and lowered by means of the global metric
tensor.

The complete action integral for the natural manifold (no external forces:
in practice no matter, in our case) will be: 
\begin{equation}
\int \left( R+\frac{1}{2}\lambda \varepsilon ^{2}+\mu \varepsilon _{\alpha
\beta }\varepsilon ^{\alpha \beta }\right) \sqrt{\left\vert g\right\vert }
dx^{N}  \label{azione}
\end{equation}

The structure of (\ref{azione}) recalls the classical form where a kinetic
and a potential terms appear. Here the role of kinetic term is played by $R$
which contains derivatives of the strain tensor.

\subsection{"Elastic" Einstein equations}

The treatment we have given of the behaviour of the "elastic-style"
space-time is not different in the form from the introduction of some
"material" contribution to the Lagrangian. Now this contribution is
expressed by (\ref{elasto}) with 
\begin{equation}
\varepsilon _{\mu \nu }=\frac{1}{2}\left( g_{\mu \nu }-\mathsf{\eta }_{\mu
\nu }\right) .  \label{diffe}
\end{equation}

One has to be careful in dealing with $\mathsf{\eta }_{\mu \nu }$. As
already said, this is a tensor on the natural manifold, but is no metric at
all for that manifold. The $\mathsf{\eta }_{\mu \nu }$ tensor can in
practice be identified with the metric of the local tangent four-dimensional
frame, comoving with the cosmic flow of the given universe. At different
cosmic times the various local $\mathsf{\eta }_{\mu \nu }$ tensors are
related to each other via boosts based on the expansion rate.

In principle from (\ref{elasto}) and using (\ref{diffe}), varying the action
integral with respect to $g_{\mu \nu }$ (or equivalently $\varepsilon _{\mu
\nu }$ which is proportional to the non trivial part of the metric tensor),
we may obtain generalized Einstein equations in the form: 
\begin{equation}
G_{\mu \nu }=T_{e\mu \nu }+\kappa T_{\mu \nu }  \label{einstein}
\end{equation}%
where all that is neither in the Einstein tensor nor in the matter
energy-momentum tensor can be interpreted as an effective "elastic" energy
momentum tensor $T_{e\mu \nu }$. Since $T_{e\mu \nu }$, as well as $T_{\mu \nu }$, is obtained varying a true
scalar (the integrand of (\ref{azione})) with respect to a true tensor, it is
also a good tensor, retaining all properties of tensors. In particular no
coordinate choice can bring $T_{e\mu \nu }$ to zero, unless it is
identically zero.

In vacuo the final matter term is absent, however now it is in general $%
R\neq 0$, provided some internal cause, like a defect, is there. The tensor $%
T_{e\mu \nu }$, though being partially built from the metric itself, plays
the role of an additional source together with the proper matter term. In
vacuo, for instance, we see that the Bianchi identities applied to the
Einstein tensor, imply that the "elastic" energy-momentum tensor is
conserved. When matter is also present, the conservation condition applies
to the sum $T_{e\mu \nu }+\kappa T_{\mu \nu }$ and, in general, the
possibility of a transfer of energy between the matter and the strain term
is given; this is not different from the mechanisms which can subtract
energy from material systems pumping it into gravitational waves in GR, but
for the fact that now the energy of the wave has a new gauge independent
interpretation. To the whole source ($T_{\mu \nu }$
and $T_{e\mu \nu }$ together) we may apply the energy conditions often
considered in GR; if the matter source is thought to be an ordinary one the
conditions are separately satisfied for it, so that we obtain constraints
for the Lam\'{e} coefficients of space-time.

Our procedure starts from an assumption of homogeneity and isotropy for
space-time. The latter constraint does not come from the specific type of
universe one wishes to implement, but from the properties of the reference
unstrained manifold. In fact, in an unstrained manifold, where neither
defects nor material sources are present, no event and no direction is
better of any other, so the system is unavoidably homogeneous and isotropic.
When considering a peculiar global configuration, induced by a defect or an
arbitrary matter distribution, the corresponding locally anisotropic strain
of the real manifold can of course induce in turn some anisotropy (and
inhomogeneity) in the elastic parameters, however, as it is the case for
ordinary three-dimensional continua, the induced anisotropy in the
properties of the "stuff" may usually be considered as second order with
respect to the strain. So, as far as the theory is assumed to be linear, the
homogeneity and isotropy of the elastic modulus tensor can be held.

\section{Specific symmetries: a Robertson Walker space-time}

\label{Specific symmetries: a Robertson Walker space-time}

\begin{figure}[tbph]
\begin{center}
\includegraphics[width=10cm]{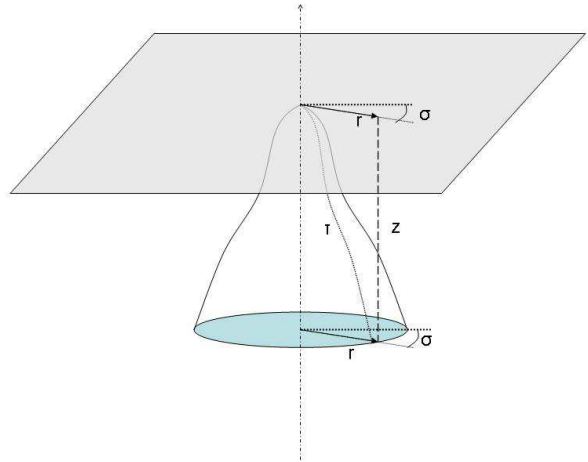}
\end{center}
\caption{A curved surface with a central symmetry is embedded in a
three-dimensional manifold. The reference frame is a plane and the global
coordinates are cylindrical.}
\label{fig.1}
\end{figure}

The purpose of the previous section was to find out an appropriate
Lagrangian for space-time and it expli\-ci\-tly depends on the strain tensor
which, in turn, depends on the symmetry. In the present work we are
interested in describing the universe as a whole so we now focus on
Robertson Walker (RW) symmetry and derive the strain tensor in such a
symmetry by means of eq. (\ref{diffe}). Willing to use the displacement
vector field (cfr. eq.(\ref{strain}) we need an embedding flat manifold.
From a geometrical point of view the RW symmetry may be interpreted as a
point symmetry so that a natural choice is to have polar coordinates around
the center of symmetry. Even though the system is four-dimensional, it is
instructive, as a first step, to describe it in two dimensions: one radial
distance, which for us will be cosmic time, and an "angle". So, in the
spirit of the previous sections, our reference manifold is a plane (with
Lorentzian signature) and the natural manifold is a curved bi-dimensional
surface (again with Lorentzian signature). Both surfaces are embedded in a
flat three-dimensional manifold and the natural choice of coordinates for it
will be the cylindrical ones. With no loss of generality we may assume that
the reference axis passes through the center of symmetry of the natural
manifold. The global configuration is reproduced in fig.\ref{fig.1}. 
The $\tau $ coordinate is the radial coordinate in the natural manifold. The
reference manifold is localized by means of the constraint $z=$ constant;
the constraint for the natural frame is $z=f\left( r\right) $ being $f$ a
non-linear function, otherwise one would have a cone, i.e. a flat
sub-manifold. The global coordinates are $r,\sigma ,z$; in the reference
frame they are $r,\sigma $ coinciding with the first pair of global
coordinates; in the natural frame the coordinates are $\tau $ and $\sigma $,
with $\sigma $ coinciding with the corresponding global coordinate. Using
the flatness of the embedding manifold we see that 
\begin{equation}
d\tau ^{2}=dr^{2}+dz^{2}=\left( 1+f^{\prime 2}\right) dr^{2}  \label{ditau}
\end{equation}%
Of course $f^{\prime }$ is a shorthand notation for $df/dr$. If $f$ is a
regular function (with the possible exception of the origin) from (\ref%
{ditau}) we can work out $r\left( \tau \right) $.

It is easy to write down the distance between two nearby points on the
reference frame directly adopting the appropriate signature: 
\begin{equation}
ds^{2}=dr^{2}-r^{2}d\sigma ^{2}+dz^{2}=dr^{2}-r^{2}d\sigma ^{2}
\label{lpiatto}
\end{equation}

The distance between the corresponding points in the natural frame is 
\begin{equation}
ds^{\prime 2}=d\tau ^{2}-r^{2}d\sigma ^{2}=\left( 1+f^{\prime 2}\right)
dr^{2}-r^{2}d\sigma ^{2}  \label{lcurvo}
\end{equation}

We can read out the metric tensor on the curved manifold from (\ref{lcurvo}
), both in the $r,\sigma $ and the $\tau ,\sigma $ coordinates, and, just to
recover the formal correspondence with the RW notation, we may identify $r$
with $a\left( \tau \right) $.

The natural coordinates for our curved manifold are $\tau $ and $\sigma $,
so, in order to proceed, we also need to cast (\ref{lpiatto}) in terms of
these coordinates; then: 
\begin{equation}
ds^{2}=\frac{d\tau ^{2}}{1+f^{\prime 2}}-a^{2}d\sigma ^{2}  \label{lpiatton}
\end{equation}

The difference between (\ref{lcurvo}) and (\ref{lpiatton}) is 
\begin{equation}
ds^{\prime 2}-ds^{2}=f^{\prime 2}dr^{2}=\frac{f^{\prime 2}}{1+f^{\prime 2}}
d\tau ^{2}  \label{differenza}
\end{equation}

Interpreting (\ref{differenza}) on the light of sect. (\ref{Review and
reformulation of the metric properties of elastic continua}) we read out the
only non-zero element of the strain tensor of the natural frame and using
the coordinates thereupon (use (\ref{ditau})) 
\begin{equation*}
\varepsilon _{\tau \tau }=\frac{f^{\prime 2}}{2\left( 1+f^{\prime 2}\right) }%
=\frac{1-\dot{a}^{2}}{2}
\end{equation*}%
With a little change in the notation we put $\dot{a}=dr/d\tau $; the reason
is to conform to the standard notation for a RW universe.

\subsection{$4+1$-dimensional embedding}

The above treatment, when applied to the full four-dimensional manifold we
use to describe the universe, actually corresponds to a negatively curved
space (space curvature constant $k=-1$), as can be seen by noting that the
choice $z=constant$ gives a flat Lorentzian ma\-ni\-fold only if $
d\sigma ^{2}=d\chi ^{2}+\sinh ^{2}\chi \left( d\theta ^{2}+\sin ^{2}\theta
d\phi ^{2}\right) $, which corresponds to a three-dimensional negatively
curved space. Of course we would like to analyze the null space curvature
case since cosmological observations point in that direction. Let us then
look a little more in detail to the geometry of our manifolds and to the
meaning of the $\sigma $ variable we used in the previous section. Actually
various embedding strategies have been adopted for similar purposes \cite%
{lachieze, rosen65}. We shall stay with our "cylindrical" symmetry approach
and write a five-dimensional flat line element in the form: 
\begin{equation}
ds_{5}^{2}=d\tau ^{2}-d\sigma _{0}^{2}+dz^{2}  \label{cinquelinea}
\end{equation}%
Here $d\sigma _{0}$ is the three-dimensional space line element in the null
space curvature case: 
\begin{equation*}
d\sigma _{0}^{2}=dr^{2}+r^{2}\left( d\theta ^{2}+\sin ^{2}\theta d\phi
^{2}\right) ;
\end{equation*}%
$z$ is the fifth coordinate of the embedding space-time. With this
coordinate choice, if we put $z=constant$ we recover a 4-dimensional
flat sub-manifold, while RW symmetry with $k=0$ is obtained with the
following transformation, as shown in \cite{lachieze, rosen65}: 
\begin{equation*}
\left\{ 
\begin{array}{rl}
r & \rightarrow a(\tau )\ r \\ 
\tau & \rightarrow \tau (r,\tau ) \\ 
z & \rightarrow z(r,\tau ),%
\end{array}%
\right.
\end{equation*}%
Here $a(t)$ is the scale factor in the four-dimensional RW metric, and the
explicit transformation for $\tau $ and $z$ are 
\begin{eqnarray*}
dt &=&\left( \dot{a}+\frac{1}{\dot{a}}+\dot{a}r^{2}\right) d\tau /2+a(\tau
)rdr \\
dz &=&\left( \dot{a}-\frac{1}{\dot{a}}-\dot{a}r^{2}\right) d\tau /2-a(\tau
)rdr.
\end{eqnarray*}%
The final purpose is to work out the strain tensor that, using (\ref{diffe}%
), turns out to only have spatial components: 
\begin{equation*}
\varepsilon _{\mu \nu }=\left\{ 
\begin{array}{cll}
0 & if & \mu =\nu =1 \\ 
\frac{1}{2}\left( 1-a^{2}\right) & if & \mu =\nu =2,3,4 \\ 
0 & if & \mu \neq \nu .%
\end{array}%
\right.
\end{equation*}%
We can now compute the trace and the second order scalar that appear in the
elastic Lagrangian (\ref{elasto}): 
\begin{equation*}
\varepsilon _{\alpha \beta }g^{\alpha \beta }=\varepsilon =\frac{3}{2}\frac{%
\left( a^{2}-1\right) }{a^{2}}\quad \quad \varepsilon _{\alpha \beta
}\varepsilon ^{\alpha \beta }=\frac{1}{3}\varepsilon ^{2}.
\end{equation*}%
In the case of non-null spatial curvature the dependence of the strain
tensor on the scale factor is different according to the sign of the
curvature parameter. In practice the embedding strategy requires different
coordinates and different transformations for each $k$ value. Recalling sec.
(\ref{Specific symmetries: a Robertson Walker space-time}), we see that the
strain tensor both for positive and negative spatial curvatures has only one
component different from zero, namely the time-time component: 
\begin{equation*}
\varepsilon _{\tau \tau }=\frac{1}{2}(1+k\dot{a}^{2}),\quad \quad k=\pm 1.
\end{equation*}%
In the case of a negative space curvature this has been explicitly worked
out in our 2+1-dimensional example. For the $k=1$ case, however, the
reference four-dimensional flat manifold has to be Euclidean, even though
the natural manifold has a Lorentzian signature. This is due to the fact
that the 3-dimensional spatial sub-manifold is a 2-sphere.

\subsection{The space-time behaviour}

Let us now maintain a RW symmetry and study the case of an empty space-time
in which space is also flat ($k=0$) as apparently it is for the actual
universe. The action integral (\ref{azione}) becomes
\begin{equation}
S=\int \left[ -6\left( a\ddot{a}+\dot{a}^{2}\right) a+L_{e}\right] d\tau ,
\label{azioneL}
\end{equation}%
with 
\begin{equation*}
L_{e}=\frac{1}{2}\left( \lambda +\frac{2}{3}\mu \right) \varepsilon ^{2}%
\sqrt{\left\vert g\right\vert }=\frac{9}{8}B\frac{\left( 1-a^{2}\right) ^{2}%
}{a},
\end{equation*}%
where $B=\lambda +\frac{2}{3}\mu $ is the bulk modulus of the continuum.
Here we are using a point Lagrangian and we shall derive the equations of
motion varying the action in eq.(\ref{azioneL}). We can remark that, in the
absence of a defect, a RW space-time would in general not be a solution of
the Lagrange equations obtained from (\ref{azione}): the symmetry (and the
defect) are a priori conditions thus forcing the action integral (\ref%
{azioneL}). 

Since second order derivatives in the Lagrangian appear linearly we can get
rid of them by means of an integration by parts so that the effective
Lagrangian density becomes 
\begin{equation}  \label{pointlagrangian}
L=6a\dot{a}^2+L_e
\end{equation}
We can work out the energy function, $W=\frac{\partial L}{\partial \dot{a}}%
\dot{a}-L$, which is, by construction, a conserved quantity: 
\begin{equation}
W=6a\dot{a}^{2}-\frac{9}{8}B\frac{\left( 1-a^{2}\right) ^{2}}{a}.
\label{energiaW}
\end{equation}
Solving (\ref{energiaW}) for $\dot{a}$ we have 
\begin{equation}
\dot{a}^{2}=\frac{W}{6a}+\frac{3}{16}B\frac{\left( 1-a^{2}\right) ^{2}}{a^{2}%
}  \label{apunto}
\end{equation}

We see that for $a\rightarrow 0$ one has $\dot{a}\rightarrow \infty $. For $%
a\rightarrow \infty $ $\dot{a}$ diverges also.

\begin{figure}[tbp]
\begin{center}
\includegraphics[width=9cm]{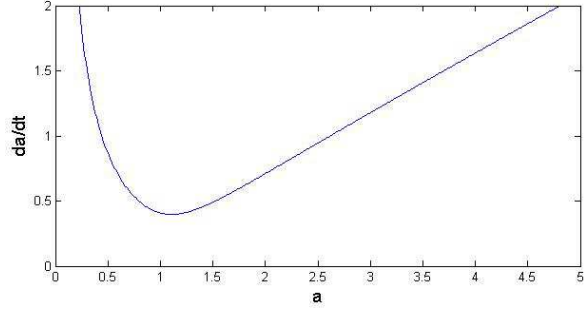}
\end{center}
\caption{Expansion rate of a Robertson Walker space-time versus the scale
factor, according to the theory. The space is assumed to be flat. Arbitrary
positive values of the parameters have been used. }
\label{fig001}
\end{figure}

The expansion rate in (\ref{apunto}) has a minimum, which means that at the
beginning the expansion is decelerated, then it becomes accelerated.\newline
The trend of (\ref{apunto}) appears in fig.(\ref{fig001}) for arbitrary
positive values of both parameters.\newline
In order to recover General Relativity in the absence of defects, the energy
function must be set to zero so that from now on we assume $W=0$.

\section{Expansion of the universe}

The main fact we want to account for is the accelerated expansion of the
universe, so in the present section we will deal with the data evidencing
this phenomenon. We of course shall proceed according to our approach, but
in the following we would also like to show how the same formulae can be
read and interpreted in more traditional ways.

\subsection{\protect\bigskip Dark fluid interpretation}

Let us put together the first integral in (\ref{apunto}) and the second
order evolution equation, deduced from the point-Lagrangian in (\ref%
{pointlagrangian}):

\begin{eqnarray}
&H^2=&\frac{3}{16}B \frac{(a^2-1)^2}{a^4} \label{fri1}\\
-2 \ddot{a}a-&\dot{a}^2=&-\frac{3}{16}B \frac{(a^2-1)}{a^2}(3a^2+1),\label{fri2}
\end{eqnarray}
$H$ is the Hubble parameter. Interpreting the r.h.s. of the equations as
representing a fluid component, we may read out the corresponding density
and pressure: 
\begin{eqnarray}
\rho&=&\frac{9}{16}B \frac{(a^2-1)^2}{a^4} \label{density}\\
p&=&-\frac{3}{16}B \frac{(a^2-1)}{a^4}(3a^2+1)\label{pressure}
\end{eqnarray}
The state parameter, i.e. $w=p/\rho $, would clearly depend on time. Since
we are interested here in late cosmology, let us derive the condition for
the acceleration to occur. An accelerating phase, $\ddot{a}>0$, requires $%
\rho +3p<0$. In our model it turns out to be 
\begin{equation}\label{acc}
\rho+3p=-\frac{9B}{8}\frac{a^4-1}{a^4},
\end{equation}
so acceleration sets in when $a^{4}>1$, or, in terms of the redshift $%
z=a_{0}/a-1$, when $z<a_{0}-1$. The parameter $a_{0}$ is the present scale
factor and its value depends on the model and the observation.\\
In particular, if we write down the equation for the "elastic" state parameter
$$
w=\frac{p}{\rho}=-\frac{1}{3}\frac{3a^2+1}{a^2-1}
$$
we can easily see that the behaviour of the elastic potential tracks radiation, curvature and cosmological constant as $a$ increases, passing from $w=1/3$ in the $a\rightarrow 0$ limit to $w=-1$ for $a\rightarrow\infty$. In a 3+1 view we may think that, close to the cosmic defect, a release of "elastic" energy in form of radiation (primordial gravitational waves) dominates; afterwards this     radiation is progressively converted into the equivalent of a "dark energy" driving an accelerated expansion, just as a cosmological constant would do.

\subsection{Type Ia supernovae luminosity}

The most direct evidence for the acceleration of the expansion comes from
the luminosity data from the type Ia supernovae. In order to test the theory
on the SnIa data we must include in our analysis the presence of matter.
This will be done as usual introducing in the Lagrangian a matter term
minimally coupled to geometry. It is 
\begin{equation}
H^2=\frac{3}{16}B\frac{\left( 1-a^{2}\right) ^{2}}{a^4}+\kappa
\sum\limits_{i}\rho _{i0}\frac{a_{0}^{3\left( 1+w_{i}\right) }}{a^{3\left(
1+w_{i}\right) }}  \label{energia}
\end{equation}

The coupling constant $\kappa $ is $16\pi G/c^{2}$ and $w_{i}$ is determined
by the equation of state of the $i_{th}$ component; $\rho _{i0}$ is the
mass/energy density of the $i_{th}$ component in the comoving frame, the
index $0$ refers to present day values. The luminosity is commonly expressed
in terms of the distance modulus and the redshift parameter appearing in the
scale factor through $a=a_{0}/\left( 1+z\right) $. It is \cite{weinberg}%
\begin{equation}
m-M=25+5\log \left( \left( 1+z\right) \int_{0}^{z}\frac{dz^{\prime }}{%
H(z^{\prime })}\right)   \label{modulus}
\end{equation}%
For (\ref{modulus}) to hold, distances have to be measured in Mpc.

The simplest is to restrict to dust and radiation, for which $w$ is
respectively $0$ and $1/3$; re-organizing the constants, the Hubble
parameter is 

\begin{equation}
H=\sqrt{\frac{B}{16}}\sqrt{ 3\left(1-\frac{(1+z)^2}{a_0^2}\right)^2+\frac{8\kappa }{3B}(1+z)^3\left[\rho_{m0}+\rho
_{r0}(1+z)\right]}  \label{hubble0}
\end{equation}
Using (\ref{hubble0}), eq.(\ref{modulus}) gives the explicit form of the distance modulus.

\subsection{Fitting the data}

Once (\ref{hubble0}) has been introduced into (\ref{modulus}) we may use an optimization
procedure in order to fit the luminosity data from type Ia supernovae. The
implied range of values of $z$ ($\leq \sim 2$) is sufficiently small to
assume that the radiation contribution is negligible ($\rho _{r0}\simeq 0$).

The distance modulus (\ref{modulus}) can be written in terms of three
optimization parameters and the new integration variable $\zeta =\left(
1+z\right) /a_{0}$ , as in the following 
\begin{equation}
m-M=\mathfrak{m}+5\log \left( \left( 1+z\right) \int_{1/a_{0}}^{\zeta }\frac{%
d\zeta ^{\prime }}{\sqrt{3\left( \zeta ^{\prime 2}-1\right) ^{2}+\psi \zeta
^{\prime 3}}}\right)   \label{formula}
\end{equation}%
where 
\begin{equation}
\left\{ 
\begin{array}{c}
\mathfrak{m}=25+\frac{5}{2}\log 16-\frac{5}{2}\log B+5\log a_{0} \\ 
\psi =\frac{8\kappa \rho _{0}}{3B}a_{0}^{3}%
\end{array}%
\right.   \label{defparametri}
\end{equation}

\begin{figure}[htbp]
\begin{center}
\includegraphics[width=9cm]{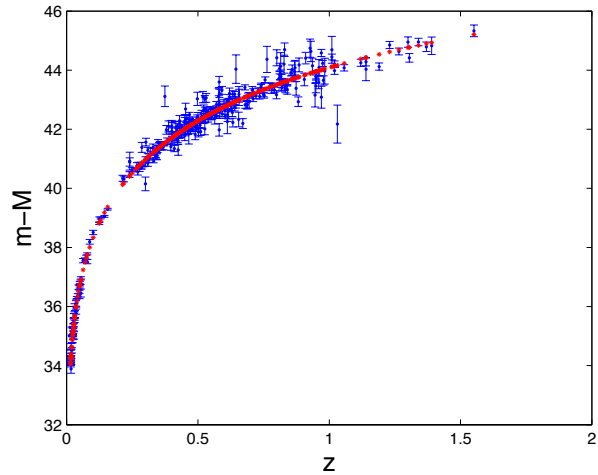}
\end{center}
\caption{Fit of the luminosity data from 307 type Ia supernovae obtained
applying the CD theory. Three optimization parameters have been used. The
reduced $\protect\chi ^{2}$ of the fit is $1.017$.}
\label{fig2}
\end{figure}

We have compared the results in (\ref{formula}) with the data of 307 SnIa's
from the Supernova Cosmology Project union survey \cite{kowalski08}. The
best fit is shown on fig.\ref{fig2}.

The optimal parameters values are:%
\begin{equation}
\left\{ 
\begin{array}{c}
\mathfrak{m}=46\pm 1 \\ 
a_{0}=1.97\pm 0.04 \\ 
\psi =10.1\pm 0.3%
\end{array}%
\right.   \label{ottimi}
\end{equation}%
Considering the different sensitivity with respect to changes in $\mathfrak{m%
}$, which is outside the logarithm in (\ref{formula}), and in $a_{0}$ and $%
\psi $ , which are in the logarithm, the optimization has been performed in
two steps. In the first step the optimization routine \cite{fminuit} has
been run with all three parameters giving the actual value for $\mathfrak{m}$
and a first estimate of $a_{0}$ and $\psi $ (actually $1.98$ and $10.18$)
with a big uncertainty (up to $100\%$ in the case of $\psi $). In the second
step $\mathfrak{m}$ has been fixed to its already found value and the
routine has been used again with the two remaining free parameters, thus
yielding a far better uncertainty. The final reduced $\chi ^{2}$ of the fit
is $1.017$.

We compare our result with what can be obtained using the $\Lambda CDM$
theory. In this case the distance modulus, expressed in terms of two free
parameters, is \cite{tartagliaaltri} 
\begin{equation}
m-M=\mu +5\log \left( 1+z\right) +5\log \int_{0}^{z}\frac{dz^{\prime }}{
\sqrt{\Omega _{m}\left( 1+z^{\prime }\right) ^{3}+1-\Omega _{m}}}
\label{milcdm}
\end{equation}
Using the same two-steps optimization process the final $\chi
^{2}$ is $1.019$, so we see that the $CD$ fit is better.

Let us now deduce some cosmological parameters from our best fit values by
using (\ref{defparametri}) and (\ref{ottimi}). Eliminating $B$ between the
first and the second equation of (\ref{defparametri}) we find a relation
among all parameters we have used for the fit, and $\rho _{0}$: 
\begin{equation}
\mathfrak{m}=25+\frac{5}{2}\log \frac{6\psi }{\kappa \rho _{0}a_{0}}.
\label{mimi}
\end{equation}%
The actual value of the mass/energy density in the universe is not easy to
assess, but is usually thought to be somewhere between $10^{-27}$ and $%
10^{-28}$ kg/m$^{3}$. \newline
By using eq.(\ref{mimi}) we can derive the estimated mass density from our
fit, and it is $\rho _{0}\sim 3.4\times 10^{-27}kg/m^{3}$. This is but a
rough estimate because the SnIa data do not allow for more accurate results.
Actually the uncertainty domain for $\rho _{0}$ is in the order of $100\%$.
Another estimate we can do concerns the value of the Hubble constant $H_{0}$%
. From (\ref{hubble0}), neglecting the contribution from radiation, we have: 
\begin{equation}\label{H0}
H_{0}=c\sqrt{\kappa \rho _{0}}\sqrt{\frac{1}{2\psi }\frac{\left(
1-a_{0}^{2}\right) ^{2}}{a_{0}}+\frac{1}{6}}
\end{equation}%
Introducing the numerical values we found for the parameters\footnote{%
Remember that for the use in the magnitude calculations all distances must
be expressed in Mpc.}, we have 
\begin{equation*}
H_{0}=64\pm 35\ \frac{km}{s\times Mpc}.
\end{equation*}%
Once more the big uncertainty comes mainly from the uncertainties in the
luminosity data of SnIa's.

Until now we have avoided the explicit use of the $B$ parameter, however
this parameter, in our theory, has the simple physical meaning of bulk
modulus of space-time, so let us compute it. From the first equation of the (%
\ref{defparametri}) we obtain: 
\begin{equation}
B=\left( 3\pm 2\right) \times 10^{-7} Mpc^{-2}=\left( 3\pm 2\right)
\times 10^{-52} m ^{-2}.  \label{bi}
\end{equation}
Looking back at equation (\ref{apunto}) we find for B the constraint $B>0$, which is indeed satisfied by (\ref{bi}).

\section{Newtonian limit of the theory}

After having found a good correspondence between theory and data at the
level of the SnIa luminosity dependence on redshift we should also verify
that a correct weak energy (Newtonian) limit exists. To that purpose we
start with some general remarks. Since in practice our theory simply
additively introduces a peculiar source term into the Lagrangian of space
time, whenever this new term is sent to zero we of course recover plain GR
with all its features and local limits. It would not be so only if the
additional term (\ref{elasto}) were somehow singular, which is not the case.
In fact, excluding the very cosmic defect, $L_{e}$ can continuously go to
zero at any place, together with the local strain.

We think this could be enough, however let us explicitly verify what the
weak field limit is. In order to make this check, let us consider a
spherically symmetric, stationary physical system. We know that the general
line element for this problem is
\begin{equation}
ds^{2}=f\left( r\right) d\tau ^{2}-h\left( r\right) dr^{2}-r^{2}d\theta
^{2}-r^{2}\sin ^{2}\theta d\varphi ^{2}  \label{sferico}
\end{equation}
where Schwarzschild coordinates are used.

It is easy to read our strain tensor (its non-zero elements) out of the
metric in (\ref{sferico}) comparing it with a Minkowski line element in
spherical coordinates. We get:
\begin{equation}
\begin{array}{cc}
\varepsilon _{00}&=\frac{f-1}{2} \\
\varepsilon _{rr}&=\frac{1-h}{2}
\end{array}
\label{strainN}
\end{equation}
Equipped with (\ref{strainN}) we are able to explicitly work out (\ref
{elasto}) for our problem, then we write the corresponding Lagrangian for
empty space-time (no matter, except at the source of the local spherical
symmetry). Modulo an unessential $\sin \theta $ factor and after eliminating
a second order derivative of $f$ by a by part integration, the explicit
formula is (primes mean derivatives with respect to $r$):
\begin{equation}
\begin{array}{cc}
L&=-\frac{2fh^{\prime }r+2fh^{2}-2fh}{h\sqrt{fh}} +\frac{\lambda }{4}\frac{\left( f-1\right) \left( h-1\right) }{\sqrt{fh}}
r^{2}\\
&+\frac{1}{4}\left( \mu +\frac{\lambda }{2}\right) \left( \frac{\left(
f-1\right) ^{2}}{f^{2}}+\frac{\left( h-1\right) ^{2}}{h^{2}}\right) \sqrt{fh}
r^{2}
\end{array}
  \label{lagru}
\end{equation}

From (\ref{lagru}) we obtain the Euler-Lagrange equations for $f$ and $h$:%

\begin{eqnarray}
fh-rf^{\prime }-f&=&\frac{r^{2}}{16fh}\left\{ \lambda \left[ \left(
\left( 2h+1\right) ^{2}-5\right) f^{2}-4fh^{2}+\left( f-h\right) ^{2}\right]\right..\nonumber\\
&&\left.+2\mu \left[ \left( 2h^{2}+2h-3\right) f^{2}+\left( 1-2f\right) h^{2}\right]
\right\} \label{eulag1}\\ 
rh^{\prime }+h^{2}-h&=&\frac{r^{2}}{16f^{2}}\left\{ \lambda \left[ \left(
4f-3\right) h^{2}+\left( 2h-1\right) ^{2}f^{2}-2fh\right]\right. \nonumber\\
&&\left.+2\mu \left[
\left( 2f+f^{2}-3\right) h^{2}+f^{2}\left( 1-h\right) ^{2}\right] \right\}
 \label{eulag2}
\end{eqnarray}

The left hand side of eq.s (\ref{eulag1}-\ref{eulag2}) corresponds to the equations for
the Schwarzschild problem, whilst the right hand side can be considered to
be "small" as far as it is
\begin{equation}
r^{2}\mu ,r^{2}\lambda <<1  \label{range}
\end{equation}

Under the latter assumption we may look for solutions like:%
\begin{equation}
\left\{ 
\begin{array}{c}
f=f_{0}+f_{1} \\ 
h=h_{0}+h_{1}
\end{array}
\right.  \label{ansatz}
\end{equation}
where 
\begin{equation*}
\left\{ 
\begin{array}{c}
f_{0}=1-2\frac{m}{r} \\ 
h_{0}=\frac{1}{f_{0}}
\end{array}
\right.
\end{equation*}
is the Schwarzschild solution ($m=GM/c^{2}$) and
\begin{equation}
f_{1},h_{1}\sim \lambda ,\mu  \label{condition}
\end{equation}

Under the ansatz (\ref{ansatz}) and condition (\ref{condition}) eq.s (\ref
{eulag1}-\ref{eulag2}) may be solved to first order in $\lambda $, $\mu $. The result is:
\begin{equation}
\left\{ 
\begin{array}{c}
f_{1}=\frac{C}{r}+\left( 3m^{2}\mu -A\right) \left( 1-2\frac{m}{r}\right)
+\lambda \left( \frac{6m^{3}+4m^{3}\ln \left( r-2m\right) }{r}-\frac{m^{4}}{%
r^{2}\left( 1-2\frac{m}{r}\right) }\right)  \\ 
+\mu \left( \frac{20m^{3}}{r}-\frac{2m^{4}}{r^{2}\left( 1-2\frac{m}{r}%
\right) }+\left( \allowbreak 22\frac{m^{3}}{r}-6m^{2}\right) \ln \left(
r-2m\right) -\frac{3}{2}mr\left( 1-2\frac{m}{r}\right) \right)  \\ 
h_{1}=\frac{C}{r\left( 1-2\frac{m}{r}\right) ^{2}}+\frac{\lambda }{r\left(
1-2\frac{m}{r}\right) ^{2}}\left( \frac{3m^{4}}{r\left( 1-2\frac{m}{r}%
\right) }-m^{2}r-4m^{3}\ln \left( r-2m\right) \right)  \\ 
+\frac{\mu }{r\left( 1-2\frac{m}{r}\right) ^{2}}\left( \frac{6m^{4}}{r\left(
1-2\frac{m}{r}\right) }-3m^{2}r-\frac{m}{2}r^{2}-10m^{3}\ln \left(
r-2m\right) \right) 
\end{array}%
\right.   \label{grosso}
\end{equation}
Here $A$ and $C$ are integration constants. $A$ is used to remove constant
contributions from $f_{1}$.

Up to this moment we have not used the assumption that the gravitational
field is weak. Let us introduce this condition now, linearizing the results
in $m/r$. The result is%
\begin{equation*}
\left\{ 
\begin{array}{c}
f_{1}\simeq \frac{C}{r}-\frac{3}{2}\mu mr \\ 
h_{1}\simeq \frac{C}{r}+4\frac{Cm}{r^{2}}-\frac{1}{2}\mu mr%
\end{array}%
\right.
\end{equation*}

As for the integration constant $C$, it contributes a redefinition of the
mass of the spherical source and a short range contribution to $h_{1}$;
fixing $C=0$ we finally have:%
\begin{equation}
\left\{ 
\begin{array}{c}
f=1-2\frac{m}{r}-\frac{3}{2}\mu mr \\ 
h=1+2\frac{m}{r}-\frac{1}{2}\mu mr%
\end{array}%
\right.   \label{Schwfinali}
\end{equation}

The solutions (\ref{Schwfinali}) are acceptable, as already said, as far as
condition (\ref{range}) holds. Now condition (\ref{range}) can be updated to 
\begin{equation*}
r<<\frac{1}{m\mu }
\end{equation*}%
If we consider the example of a Sun-like star, where it is $m\sim 10^{3}$ m,
and use the result found for $B$ in the previous section (reasonably it is
also $\mu \sim B$) we obtain%
\begin{equation*}
r<<10^{49} m \sim 10^{27} Mpc \sim 10^{33} light years
\end{equation*}

As we see all deviations of our theory from the standard GR, at the Solar
System or galaxy clusters level, are absolutely negligible. The only effect
is at the cosmic level.

\section{Conclusion}

We have exploited the existing analogy between the theory of elasticity and
GR and this approach has given good fruits, however, recalling the open
questions we posed in the Introduction, we remark here that an analogy is
not an identity; we are not allowed to mechanically transpose one theory on
top of the other. In GR one properly looks for "static" solutions in four
dimensions: various possible configurations of the full four-dimensional
universe are studied, but there is no evolution because there is no
evolution parameter out of the manifold. Time is part of the manifold and
the dynamic term in the Lagrangian is the scalar curvature which contains
second order derivatives of the Lagrangian coordinates, i.e. of the elements
of the metric tensor, with respect to an arbitrary set of Gaussian
coordinates. In the case of elasticity the three-dimensional manifold has
Euclidean signature and time is the absolute Newtonian time; the dynamics is
expressed by time derivatives. The difference is paramount.

The role of the defect(s) in our theory deserves also some additional
comments. There are indeed, in the three-dimensional theory of continua,
situations where one has to do with continuous distributions of
(micro)defects, rather than with localized defects. This happens mainly as a
consequence of a plastic deformation: an initial stress in the material is
eased (for instance by reheating) and produces a distributed defects field.
The final result of this process is indeed the disappearance of the internal
strain and a permanent plastic deformation of the manifold. This situation
does not correspond to what we want to describe, since for us the strain
represents the gravitational field, or the non-trivial part of the metric
tensor,\emph{\ }and we want it to stay there and disappear only when its
causes are removed (no plastic shear). This is the reason why our attention
is focused on a localized defect responsible for a global spontaneous strain
state and for the related symmetry. An appealing possibility could be to
have a plurality of localized defects which would give rise to some more
complicated strain pattern; one could be tempted to identify these defects
with matter. We have not dared, for the moment, to pursue this idea, so in
our theory, at present, matter is treated as usually it is, appearing in the
Lagrangian as an additional independent term coupled with the (strained) geometry
via the metric tensor.

In the above conceptual background  we have combined the classical GR approach with the description of
space-time by means of the linear theory of elasticity, preserving general
covariance and all the features of GR. The expansion of the universe is
consequently described in terms of a strained four-dimensional continuum
(space-time), whose strain has partly an intrinsic origin due to the
presence of a cosmic defect, partly depends on extrinsic sources, i.e.
matter fields. The cosmic defect fixes the global symmetry of the universe,
then the general features of what, in our ($3+1$)-split view, is the cosmic
expansion. This approach can be applied to any kind of universe with any
global symmetry depending on the possible defects, just as it happens in
three dimensional solid materials. We have then assumed the basic "stuff",
i.e. space-time, to be locally homogeneous and isotropic, as a consequence
of the global homogeneity and isotropy of the flat unstrained reference
manifold and of the linearity of the theory. Working out the global
configuration of space-time for a RW symmetry (the typical symmetry assumed
to hold, on a cosmic scale, for our universe) we have found that it
naturally includes an initial extremely rapid expansion with a steeply
decreasing expansion rate, followed by acceleration. We have also verified
that the inclusion of matter in the form of fluid(s) preserving the global
symmetry does not modify the general structure of the expansion. The theory
depends on three parameters, which are the present scale factor of the
universe $a_{0}$, the bulk modulus of space-time $B$, and the present day
matter/energy density of the universe $\rho _{0}$. Using these three
quantities as optimization parameters we have fitted the luminosity data of
type Ia supernovae. The result has been good, and the value obtained for $%
\rho _{0}$ is consistent with the current estimates for barionic matter
without a need for more matter, but the uncertainty due to the accuracy
of the luminosity data is very high. According to the theory no further dark
energy is needed; however, if we wish, we may read our elastic contribution as a dark energy fluid,
whose density and pressure have been explicitly written, it would however be
rather difficult to find a reasonable physical interpretation for the
properties of this peculiar fluid.

A final remark concerns the signature of our manifolds. It always is
Lorentzian in the natural manifold which we want to represent the actual
universe. As for the reference manifold it can either be Euclidean or
Minkowskian and the embedding strategy can easily produce one or the other
of them. It is sufficient to assume the embedding flat higher dimensional
manifold to be Minkowskian. Whenever then the reference submanifold is a
space-like hyperplane (time-like normal vector) its geometry is naturally
Euclidean; viceversa, choosing as a reference submanifold a time-like
hyperplane, it will turn out to be Minkowskian.

Although the theory has been applied to the cosmic scale, we have also
verified that it has a correct Newtonian limit and, with the value of the
parameters obtained from the SnIa's fit, it is indistinguishable from GR at
the Solar system as well as at the galaxy clusters scale.

\end{document}